\documentstyle[12pt,thmsa,sw20aip]{article}
\input tcilatex
\begin{document}

\title{Perturbations of rotating cosmological black holes}
\author{Zolt{\'{a}}n Perj{\'{e}}s \\
%EndAName
KFKI Research Institute for Particle and Nuclear Physics,\\
H--1525, Budapest 114, P.O.B.\ 49, Hungary\\
{\tt e-mail: perjes@rmki.kfki.hu}}
\date{\today }
\maketitle

\begin{abstract}
Charged, rotating black hole solutions of Einstein's gravitational equations
are investigated in the presence of a cosmological constant. A pair of wave
equations governing the electromagnetic and gravitational perturbations are
derived.

\end{abstract}

\section{Introduction}

Recent observations\cite{Obs} support the presence of a cosmological term in
the gravitational equations of Einstein. If this is the case then an exact
description of cosmological black holes will not be by a solution of the
short Einstein equations but by a solution of the equations including the
cosmological constant. Indeed, the most general stationary black hole state
due to Carter does contain the cosmological constant.

The Carter space-time with cosmological constant $\Lambda $, mass ${\rm m}$,
rotation parameter $a$ and electric charge ${\rm e}$ has the metric\cite
{Carter} 
\begin{eqnarray}
ds^2 &=&\frac 1{(1+\alpha )^2\zeta \overline{\zeta }}\left\{ {\bf \Delta }%
(du-a\sin ^2\vartheta d\phi )^2-\Delta _\vartheta \sin ^2\vartheta \left[
adu-(r^2+a^2)d\phi \right] ^2\right\}  \nonumber  \label{ds2} \\
&&-\zeta \overline{\zeta }\left( \frac{dr^2}{{\bf \Delta }}+\frac{d\vartheta
^2}{\Delta _\vartheta }\right)  \label{ds2}
\end{eqnarray}
where 
\begin{equation}
{\bf \Delta }=\left( r^2+a^2\right) \left( 1-\alpha \tfrac{r^2}{a^2}\right)
-2{\rm m}r+{\rm e}^2,\qquad \qquad \Delta _\vartheta =1+\alpha \cos
^2\vartheta
\end{equation}
and 
\begin{equation}
\alpha =\frac{\Lambda a^2}3,\qquad \qquad \zeta =r-ia\cos \vartheta .
\end{equation}

In the special case when the mass ${\rm m}$, the electric charge ${\rm e}$
and the rotation parameter $a$ all vanish, the metric (\ref{ds2}) describes
the de Sitter space time, to which the metric tends for large values of $r.$
Another limit is that with a vanishing cosmological constant $\Lambda $, in
which case the metric (\ref{ds2}) describes the Kerr-Newman space-time.

In the generic case, the condition ${\bf \Delta }=0$ has up to four distinct
roots. The number of roots clearly depends on the values of the parameters.
The equilibrium solutions have charge ${\rm e}$ equal to mass ${\rm m}$ up
to an overall sign. For a positive cosmological constant, $\Lambda >0,$ this
is no longer the condition of extremality, as the equation $\left( r-{\rm m}%
\right) ^2+a^2-\alpha r^2\left( \tfrac{r^2}{a^2}+1\right) =0$ then has real
roots.

The main purpose of the present paper is to present the equations governing
the coupled electromagnetic and gravitational perturbations of rotating
cosmological black holes. These equations are separable and they generalize
the master equations for the perturbations of Kerr-Newman black holes\cite
{Perjes}.

\section{\protect\bigskip Analytic extension}

The metric (\ref{ds2}) is singular on the horizon hypersurface given by the
largest root of ${\bf \Delta }=0{\bf .}$ One can transform to a better
coordinate system by

\begin{eqnarray}
dt &=&du+(1+\alpha )\tfrac{r^2+a^2}{{\bf \Delta }}dr  \label{nulltr} \\
d\varphi &=&d\phi +(1+\alpha )\tfrac a{{\bf \Delta }}dr  \nonumber
\end{eqnarray}
which is a generalization of the inverse Boyer-Lindquist transformation. In
the new coordinates, the metric acquires the null form

\begin{eqnarray}
ds^2 &=&\frac 1{\left( 1+\alpha \right) ^2}\left[ 1-\tfrac{2{\rm m}r-{\rm e}%
^2}{\zeta \overline{\zeta }}-\alpha (\tfrac{r^2}{a^2}+\sin ^2\vartheta
)\right] \left( dt-a\sin ^2\vartheta d\varphi \right) ^2  \nonumber
\label{ds22} \\
&&+\frac 2{1+\alpha }\left( dt-a\sin ^2\vartheta d\varphi \right) \left( dr+%
\frac{\Delta _\vartheta }{1+\alpha }a\sin ^2\vartheta d\varphi \right)
\label{ds22} \\
&&-\zeta \overline{\zeta }\left( \frac{d\vartheta ^2}{\Delta _\vartheta }+%
\frac{\Delta _\vartheta }{\left( 1+\alpha \right) ^2}\sin ^2\vartheta
d\varphi ^2\right)  \nonumber
\end{eqnarray}
with the four-potential 
\begin{equation}
A=-\tfrac{{\rm e}r}{\left( 1+\alpha \right) ^2\zeta \overline{\zeta }}\left(
dt-a\sin ^2\vartheta d\varphi \right) .  \label{A}
\end{equation}
The coordinates $t$ and $r$ run from $-\infty $ to $\infty $ while $%
\vartheta $ and $\varphi $ are coordinates on a distorted 2-sphere such that 
$\varphi $ is periodic with period $2\pi $ and $\vartheta $ ranges from $0$
to $\pi $.

The singularities are conveniently explored by introducing the null tetrad%
\cite{NP} 
\begin{eqnarray}
D &\equiv &\ell ^a\tfrac \partial {\partial x^a}=\tfrac \partial {\partial r}
\nonumber \\
\Delta &\equiv &n^a\tfrac \partial {\partial x^a}=\tfrac 12\left( 1+\alpha
\right) ^2\left[ \tfrac{2{\rm m}r-{\rm e}^2}{\zeta \overline{\zeta }}%
-1+\alpha (\tfrac{r^2}{a^2}+\sin ^2\vartheta )\right] \tfrac \partial
{\partial r}+(1+\alpha )\tfrac \partial {\partial t} \\
\delta &\equiv &m^a\tfrac \partial {\partial x^a}=\tfrac 1{2^{1/2}\overline{%
\zeta }}\left[ \Delta _\vartheta ^{1/2}\left( \tfrac \partial {\partial
\vartheta }-ia\sin \vartheta \tfrac \partial {\partial r}\right) +\tfrac{%
1+\alpha }{\Delta _\vartheta ^{1/2}}\left( \tfrac i{\sin \vartheta }\tfrac
\partial {\partial \varphi }+ia\sin \vartheta \tfrac \partial {\partial
t}\right) \right]  \nonumber \\
\overline{\delta } &\equiv &\overline{m}^a\tfrac \partial {\partial x^a}. 
\nonumber
\end{eqnarray}
In the NP notation, the Maxwell tensor components are 
\begin{eqnarray}
\Phi _0 &\equiv &F_{ab}\ell ^am^b=0  \nonumber  \label{Phis} \\
\Phi _1 &\equiv &\tfrac 12F_{ab}\left( \ell ^an^b+\overline{m}^am^b\right) =%
\frac{{\rm e}}{2^{1/2}\zeta ^2}  \label{Phis} \\
\Phi _2 &\equiv &F_{ab}\overline{m}^an^b=\frac{i{\rm e}a\sin \vartheta }{%
\zeta ^3}\Delta _\vartheta ^{1/2}.  \nonumber
\end{eqnarray}
The significant components of the Weyl curvature $C_{abcd}$ are 
\begin{eqnarray}
\Psi _2 &=&\frac{{\rm e}^2-{\rm m}\overline{\zeta }}{\zeta ^3\overline{\zeta 
}}  \nonumber \\
\Psi _3 &=&-3i\Delta _\vartheta ^{1/2}a\sin \vartheta \frac{{\rm m}\overline{%
\zeta }-{\rm e}^2}{2^{1/2}\zeta ^4\overline{\zeta }} \\
\Psi _4 &=&3\Delta _\vartheta a^2\sin ^2\vartheta \frac{{\rm m}\overline{%
\zeta }-{\rm e}^2}{\zeta ^5\overline{\zeta }}  \nonumber
\end{eqnarray}
and the remaining two components vanish, $\Psi _0\equiv -C_{abcd}\ell
^am^b\ell ^cm^d=0$ and $\Psi _1\equiv -C_{abcd}\ell ^an^b\ell ^cm^d=0$. The
curvature singularities are at $r=0$ and $\vartheta =\pi /2$ .

The null coordinates $\left( r,\vartheta ,t,\varphi \right) $ cover a part
of the manifold extending beyond the hypersurface ${\bf \Delta }=0$ . A
different extension is given by the twin transformation

\begin{eqnarray}
d\widetilde{t} &=&du-(1+\alpha )\tfrac{r^2+a^2}{{\bf \Delta }}dr
\label{tildetr} \\
d\widetilde{\varphi } &=&d\phi -(1+\alpha )\tfrac a{{\bf \Delta }}dr. 
\nonumber
\end{eqnarray}
The analytic extension of the space-time can be traversed by transiting
directly from the coordinate patches $\left( r,\vartheta ,t,\varphi \right) $
to $\left( r,\vartheta ,\widetilde{t},\widetilde{\varphi }\right) $ by use
of the combined transformations (\ref{nulltr}) and (\ref{tildetr}) on each
of the domains of the space-time lying in between the zeroes of the function 
$\Delta .$ The complete analytic extension of the uncharged Kerr-de Sitter
space-time has been briefly discussed in Ref. 4. There is a one-to-one
correspondence between the structure of the charged and the uncharged
metrics. The diagrammatic representation on Fig. 6.5 in Ref. 4 of the
complete manifold remains valid for the charged Carter space-time.

\section{Black hole perturbations}

We choose the perturbed tetrad such that the spinor $o^A$ is one of the four
principal spinors of the Weyl curvature: 
\begin{equation}
\Psi _0=0.  \label{pnd}
\end{equation}
The quantity $\Psi _0$ transforms under the infinitesimal dyad
transformation 
\begin{equation}
o^A\rightarrow o^A+b\iota ^A,\qquad \iota ^A\rightarrow \iota ^A,
\label{inftr}
\end{equation}
as follows: 
\begin{equation}
\Psi _0\rightarrow \Psi _0+4b\Psi _1.
\end{equation}
Here $b$ is an arbitrary but small complex multiplier function such that
higher powers of $b$ are negligible. Since the curvature quantity $\Psi _1$
itself is small, the spinor $o^A$ remains a principal spinor of the
curvature under the transformations (\ref{inftr}). We use this gauge freedom
to eliminate all coupling terms from the wave equation 
\begin{equation}
\Box _1\phi =0  \label{Phieq}
\end{equation}
where the wave operator is 
\begin{eqnarray}
\Box _s &=&{\bf \Delta }^{-s}\tfrac \partial {\partial r}{\bf \Delta }%
^{s+1}\tfrac \partial {\partial r}+\tfrac 1{\sin \vartheta }\tfrac \partial
{\partial \vartheta }\Delta _\vartheta \sin \vartheta \tfrac \partial
{\partial \vartheta } \\
&&+\tfrac{1+\alpha }{\Delta _\vartheta \sin ^2\vartheta }\left[ \left(
1+\alpha \right) \tfrac \partial {\partial \varphi }+2is\left( \Delta
_\vartheta -\alpha \sin ^2\vartheta \right) \cos \vartheta \right] \tfrac
\partial {\partial \varphi }  \nonumber \\
&&{\cal +}2\left( 1+\alpha \right) a\left( \tfrac{1+\alpha }{\Delta
_\vartheta }\tfrac \partial {\partial t}-\tfrac \partial {\partial r}\right)
\tfrac \partial {\partial \varphi }  \nonumber \\
&&+\tfrac{\left( 1+\alpha \right) ^2}{\Delta _\vartheta }a^2\sin ^2\vartheta 
\tfrac{\partial ^2}{\partial t^2}-2\left( 1+\alpha \right) (r^2+a^2)\tfrac{%
\partial ^2}{\partial r\partial t}  \nonumber \\
&&-2\left( 1+\alpha \right) \left[ (s+2)r+\left( \Delta _\vartheta +s\alpha
\sin ^2\vartheta \right) ia\cos \vartheta \right] \tfrac \partial {\partial
t}  \nonumber \\
&&-12\tfrac{\alpha s}{a^2}r^2+\left( 1-\alpha -6\alpha \cos ^2\vartheta
\right) s  \nonumber \\
&&-\tfrac{s^2\cos ^2\vartheta }{\sin ^2\vartheta }\Delta _\vartheta \left(
1-\tfrac \alpha {\Delta _\vartheta }\sin ^2\vartheta \right) ^2,  \nonumber
\end{eqnarray}
and the electromagnetic perturbation function $\phi $ is defined by 
\begin{equation}
\Phi _0={\rm e}\phi .  \label{phidef}
\end{equation}
%The vanishing of the coupling terms yields the intrinsic derivative of
%the first curvature:
%\begin{equation}
%\Delta \kappa =\left( 3\gamma -\overline{\mu }+\overline{\gamma
%}-\frac{%
%\Delta \Phi _1}{\Phi _1}\right) \kappa +\left( \overline{\delta
%}-4\alpha +%
%\frac{\overline{\delta }\Phi _1}{\Phi _1}\right) \sigma -2\Psi _1.
%\label{Delka}
%\end{equation}

The second wave equation can be obtained by essentially repeating the steps
made in Ref. 6 for deriving the master equation of the gravitational
perturbation components. %An additional intrinsic
%derivative of the first curvature is obtained by acting with the
%commutator $[\delta,D]$ on the function $\Psi_1$:
In the resulting equation, each term on the left contains the small function 
\begin{equation}
\psi =\frac \sigma {\zeta ^2}  \label{psidef}
\end{equation}
and each term on the right contains either a $\phi $ or the first-order spin
coefficient $\kappa \equiv \ell _{a;b}m^a\ell ^b$. When inserting the
unperturbed values of the operators and factors, neither the $\psi $ terms,
nor the $\phi $ terms cancel but those with $\kappa $ do. We get the wave
equation 
\begin{equation}
\Box _2\psi =\frac 1{\overline{\zeta }^2}J\phi .  \label{psieq}
\end{equation}
Thus the solution $\phi $ of Eq. (\ref{Phieq}) will provide the source
function $J\phi $ for the equation for $\psi .$ The source term is a
functional of the field $\phi $ containing up to second derivatives, with
the operator

\begin{eqnarray}
J &=&\left[ ({\rm e}^2-{\rm m}\zeta )+\tfrac \alpha {a^2}\zeta \bar{\zeta}%
^3\right]  \nonumber \\
&&\times \left[ \Delta _\vartheta ^{1/2}\left( \tfrac \partial {\partial
\vartheta }-\tfrac{\cos \vartheta }{\sin \vartheta }\right) +i\tfrac{%
1+\alpha }{\Delta _\vartheta ^{1/2}}\left( a\sin \vartheta \tfrac \partial
{\partial t}+\tfrac 1{\sin \vartheta }\tfrac \partial {\partial \varphi
}\right) +\tfrac \alpha {\Delta _\vartheta ^{1/2}}\sin \vartheta \cos
\vartheta \right] \tfrac \partial {\partial r}  \nonumber \\
&&-\tfrac 1{\bar{\zeta}}\left[ {\rm e}^2+\left( {\rm m}-\tfrac \alpha {a^2}%
\bar{\zeta}^3\right) (2\bar{\zeta}-\zeta )\right] \\
&&\times \left[ \Delta _\vartheta ^{1/2}\left( \tfrac \partial {\partial
\vartheta }-\tfrac{\cos \vartheta }{\sin \vartheta }\right) +i\tfrac{%
1+\alpha }{\Delta _\vartheta ^{1/2}}\left( a\sin \vartheta \tfrac \partial
{\partial t}+\tfrac 1{\sin \vartheta }\tfrac \partial {\partial \varphi
}\right) +\tfrac \alpha {\Delta _\vartheta ^{1/2}}\sin \vartheta \cos
\vartheta \right]  \nonumber \\
&&+\tfrac 1{\bar{\zeta}}\left[ {\rm e}^2-\left( {\rm m}-\tfrac \alpha {a^2}%
\bar{\zeta}^3\right) (\zeta +2\bar{\zeta})\right] ia\sin \vartheta \Delta
_\vartheta ^{1/2}\tfrac \partial {\partial t}.  \nonumber
\end{eqnarray}
One can treat these equations by expanding the homogeneous solutions in
quasi-normal modes with energy $\omega $ and helicity\cite{Teukolsky} $m$: 
\begin{equation}
\Phi _0=\int d\omega \sum_{l,m}R\left( r\right) S_l^m\left( \vartheta
\right) e^{i(m\varphi -\omega t)}.
\end{equation}
Here the radial function $R\left( r\right) $ and the angular function $%
S_l^m\left( \vartheta \right) $ satisfy the ordinary differential equations,
respectively, 
\begin{eqnarray}
&&\left\{ {\bf \Delta }^{-s}\tfrac \partial {\partial r}{\bf \Delta }%
^{s+1}\tfrac \partial {\partial r}+2i\left( 1+\alpha \right) \left[
(r^2+a^2)\omega -am\right] \tfrac \partial {\partial r}\right.  \nonumber \\
&&\left. +2i\omega \left( 1+\alpha \right) (s+2)r-12\tfrac{\alpha s}{a^2}%
r^2-K\right\} R=0  \label{radeq}
\end{eqnarray}
\newpage
\begin{eqnarray}
&&\left\{ \tfrac 1{\sin \vartheta }\tfrac \partial {\partial \vartheta
}\Delta _\vartheta \sin \vartheta \tfrac \partial {\partial \vartheta
}\right.  \nonumber  \label{angeq} \\
&&{\cal -}\tfrac 1{\Delta _\vartheta }\left[ \tfrac{\left( 1+\alpha \right)
m+s\Delta _\vartheta \cos \vartheta }{\sin \vartheta }-\left( 1+\alpha
\right) \omega a\sin \vartheta -\alpha s\cos \vartheta \sin \vartheta
\right] ^2  \label{angeq} \\
&&\left. +\left( 1-\alpha -6\alpha \cos ^2\vartheta \right) s-2\left(
1+\alpha \right) \left( s+1\right) \omega a\cos \vartheta +K\right\} S_l^m=0
\nonumber  \label{angeq}
\end{eqnarray}
and $K$ is the separation constant of the kernel of operator $\Box _s$.

In the present coordinate system, the metric (\ref{ds2}) remains regular on
the null hypersurfaces ${\bf \Delta }=0.$ The radial equation (\ref{radeq})
has a singularity on the horizon (the outer solution of ${\bf \Delta }=0$),
but, as was shown in Ref. 8, one can choose the boundary conditions for the
wave equations (\ref{radeq}) and (\ref{angeq}) on the horizon such that the
perturbations $\phi $ and $\psi $ are regular.

In terms of a solution of Eqs. (\ref{Phieq}) and (\ref{psieq}), the
perturbation functions $\sigma $ and $\Phi _0$ are available from the
respective simple relations (\ref{psidef}) and (\ref{phidef}). One can next
compute the spin coefficient $\kappa $ by a method quite similar to that
described in Ref. 6.

\section{Conclusions}

Equations (\ref{Phieq}) and (\ref{psieq}) generalize the perturbation
equations of Ref. 6 for black holes in the presence of a cosmological
constant. The nature of these equations is quite unprecedented in general
relativity. Their new feature is that they refer to the perturbations of a
well-defined geometrical object in an electrovacuum (or vacuum) space-time.
This object is a principal null congruence. In contrast, the first known
master equation for the Kerr black hole\cite{Teukolsky} -- also a linear
hyperbolic equation -- describes the perturbations of the component $\Psi _0$
of the Weyl curvature. The latter quantity is {\it not} uniquely defined,
since it changes with tetrad rotations unless those are restricted to (\ref
{inftr}).

\section*{Acknowledgments}

I thank M\'{a}ty\'{a}s Vas\'{u}th for discussions. This work has been
supported by the OTKA grant T031724.


\begin{thebibliography}{9}
\bibitem{Obs}  B. Schmidt et al., Astrophys. J. {\bf 507}, 46 (1998), S.
Perlmutter et al., Astrophys. J. {\bf 517}, 565 (1999), P. de Bernardis et
al., Nature {\bf 404}, 995 (2000)

\bibitem{Carter}  B. Carter, Commun. Math. Phys. {\bf 16}, 280 (1968)

\bibitem{NP}  E. T. Newman and R. Penrose, J. Math. Phys. {\bf 3}, 566
(1962), to be referred to as NP.

\bibitem{Carter3}  B. Carter, in {\it Black Holes}, Eds. C. DeWitt and B. S.
DeWitt, Gordon and Breach, 1972

\bibitem{Carter2}  B. Carter, Phys. Rev. {\bf 174}, 1559 (1969)

\bibitem{Perjes}  Z. Perj\'{e}s, gr-qc/0206088 (2002)

\bibitem{Teukolsky}  S. A. Teukolsky, Phys. Rev. Letters {\bf 29}, 1114
(1972)

\bibitem{bhpart}  Z. Perj\'{e}s and M. Vas\'{u}th, to appear in Ap.J.
\end{thebibliography}
\end{document}